\documentclass[twocolumn,preprintnumbers,amsmath,amssymb,prb]{revtex4}
\usepackage{graphicx}
\usepackage{dcolumn}
\usepackage[usenames,dvipsnames]{color}
\usepackage{bm}

\newcommand \comma {\mbox{\makebox[.1 in]{ },}}

\def \be {\begin{equation}}
\def \ee {\end{equation}}
\def \ben {\begin{eqnarray}}
\def \een {\end{eqnarray}}

\begin{document}
\bibliographystyle{jpc}
\title{Is There Elliptic Distortion in the Light Harvesting Complex 2 of Purple Bacteria?}

\author{Seogjoo Jang}
\affiliation{Department of Chemistry and Biochemistry, Queens College of the City University of New York, 65-30 Kissena Boulevard, Flushing, New York 11367-1597}

\author{Robert J. Silbey}
\affiliation{Department of Chemistry, Massachusetts Institute of Technology, Cambridge, Massachusetts 02139}

\author{Ralf Kunz, Clemens Hofmann, and J\"{u}rgen K\"{o}hler}
\affiliation{Experimental Physics IV and Bayreuth Institute of Macromolecular Research (BIMF), Universit\"{a}t Bayreuth, 95447 Bayreuth, Germany}
\date{Published in the {\it Journal of Physical Chemistry B} {\bf 115}, 12947-12953 (2011) }

\begin{abstract}
Single molecule spectroscopy (SMS) revealed unusually large gap between two major exciton peaks of the B850 unit of light harvesting complex 2 (LH2), which could be explained assuming elliptic distortion or $k=2$ symmetry modulation in the site excitation energy.  On the basis of extensive simulation of the SMS data and ensemble lineshape, we found that uniform modulation of $k=2$ symmetry cannot explain the dependence of intensity ratios on the gap of the two major peaks, which are available from SMS, nor the ensemble lineshape.  Alternative models of disorder with $k=1$ and $k=2$ symmetry correlation are shown to reproduce these data reasonably well and can even explain the gap distribution when it is assumed that the lower major peak in the SMS lineshape is an intensity weighted average of $k=1-$ and $k=0$ states.

\end{abstract}

\maketitle

\noindent

\section{Introduction}
How plants harvest photons from sunlight is a question that continues motivating various spectroscopic and theoretical studies. One of the most highly studied systems in this quest is the light harvesting complex 2 (LH2) of purple bacteria, for which the determination of X-ray structure in 1995\cite{mcdermott-nature374,koepke-structure4} has generated a surge of experimental and theoretical investigations.\cite{sundstrom-jpcb103,hu-qrb35}   These studies provided convincing evidence for substantial delocalization of the exciton of the B850 unit  in LH2.   However, direct experimental confirmation of such delocalized exciton state had remained difficult to get. The single molecule spectroscopy (SMS)\cite{vanojien-science285,cogdell-qrb39,cogdell-bcj422} of LH2 by van Oijen {\it et al.} was a significant breakthrough in this sense because it clearly demonstrated distinctive excitonic peaks characteristic of delocalized exciton states for the B850 unit.  

While most of the SMS data\cite{vanojien-science285} could be understood by the Frenkel exciton model based on the X-ray crystal structure and simple Gaussian site energy disorder, one feature -- the large gap between two major peaks in each SMS lineshape of the B850 unit -- was not easy to explain.  For this, a new structural model of LH2 with elliptic distortion was proposed.\cite{vanojien-science285} Since then, follow-up justifications\cite{matsushita-bj80} and relevant theoretical analyses\cite{mostovoy-jpcb104,dempster-jcp114} were made.  Most recently, an improved model that assumes elliptic modulation in the site energy\cite{hofmann-cpl395} rather than in the structure was developed. 
While this latest model seems to provide satisfactory description of most SMS data, its test against other ensemble/subensemble spectroscopic data has not been made yet.\cite{sundstrom-jpcb103,scholes-jpcb104,matsuzaki-jpcb105} 

Due to the fragile nature of Frenkel exciton and the complexity of the biological system, it 
is possible that fine details of structural and disorder models may vary with each spectroscopy and sample.  On the other hand, because of the fact that the disorder in LH2 is of intermediate magnitude at most, such differences in general do not cause significant qualitative changes in many spectroscopic data.   In this sense,  the large gap distribution observed from SMS is distinct,  and more comprehensive theoretical examination of its implication is needed.  In this examination, the following two questions arise naturally: (1) Is the sample of LH2's used for SMS different from those used in typical ensemble spectroscopy?  (2) If ensemble spectroscopy is conducted for the same sample of SMS, can it  be also explained by the same model of elliptic distortion?  We intend to provide more definite answers for these questions as well.  

On the basis of the exciton Hamiltonian and exciton-bath Hamiltonian developed previously,  we here conduct comprehensive modeling of SMS data and ensemble lineshapes.   For the latter, the 2nd order time-nonlocal Quantum Master Equation (QME) is used.   Four representative models of disorder are tested in detail.  Both SMS data and the ensemble lineshape taken for the sample are used in order to check the consistency of the model.  The implications of the present study are provided in the Concluding  Remarks.

\section{Implications of the results of SMS data} 

According to the X-ray crystallography,\cite{mcdermott-nature374,koepke-structure4} LH2 is a cylindrically shaped protein-chromophore complex consisting of two groups of circularly arranged bacteriochlorophylls (BChl's) with 8 or 9 fold symmetry.  These two circular modules absorb photons at about 800 nm and 850 nm at room temperature, and are thus named B800 and B850.  A great deal of structural and spectroscopic information on  these units is available now.\cite{sundstrom-jpcb103,hu-qrb35,cogdell-qrb39} The LH2  of {\it Rhodopseudomonas acidophila} studied by van Oijen {\it et al.}\cite{vanojien-science285} has  9 fold symmetry, where the B850 unit consists of 9 pairs of BChls attached to $\alpha$ and $\beta$ polypeptides.  In the absence of disorder, the Hamiltonian representing single exciton states of B850 is given by 
\be
H_e^0=\sum_{n,m=1}^{9}\sum_{s,s'=\alpha}^\beta J_{ss'}(n-m)|s_n\rangle \langle s'_m|\comma \label{eq:h_0}
\ee
where $J_{\alpha\alpha}(0)=\epsilon_\alpha$ and $J_{\beta\beta}(0)=\epsilon_\beta$,  
the excitation energies of $\alpha$ and $\beta$ BChls.  Otherwise $J_{ss'}(n-m)$ represents the 
resonance coupling between site excitation states $|s_n\rangle$ and $|s'_m\rangle$.   

In reality, there are disorder and defects in the protein-chromophore complex, which result in an additional disorder term, $\delta H_e$. Thus, the total electronic Hamiltonian becomes 
\be 
H_e=H_e^0+\delta H_e\ . \label{eq:h_e}
\ee
Though significant, the matrix elements of $\delta H_e$ are comparable in magnitude to the off-diagonal elements of $H_e^0$. In this case,\cite{jang-jpcb105} the eigenstates of $H_e$ retain the major double band structure of $H_e^0$, and can be classified as  $|\psi_{l,p}\rangle$ and  $|\psi_{u,p}\rangle$ with $p=0,\cdots,8$, where $l$ ($u$) represents the lower (upper) band, as follows:
\be
H_e=\sum_{p=0}^{8}\left \{ {\mathcal E}_{l,p}|\psi_{l,p}\rangle \langle \psi_{l,p}|+{\mathcal E}_{u,p} |\psi_{u,p}\rangle \langle \psi_{u,p}|\right\}\ . \label{eq:he2}
\ee
In the above equation, the convention is that ${\mathcal E}_{l,p} < {\mathcal E}_{l,p'}$  
and ${\mathcal E}_{u,p} > {\mathcal E}_{u,p'}$ for $p <p'$.   
In the limit where $\delta H_e\rightarrow 0$,  each of these eigenstates reduces to that of $H_e^0$ with corresponding exciton wave number  $k=\pm(p+1)/2$, where integer division is assumed.  For example, $|\psi_{l,0}\rangle$  
reduces to $k=0$ state of the $l$ band of $H_e^0$, while 
$|\psi_{l,1} \rangle$ and  $|\psi_{l,2}\rangle $ reduce to linear combinations of 
$k=\pm 1$ states.  As will be described in more detail below, the analysis is focused on these  three lowest exciton states, which will be hereafter abbreviated as $|0\rangle$, $|1-\rangle$, and $|1+\rangle$.     \vspace{.1in}\\

 \begin{figure}
\includegraphics[width=3.2 in]{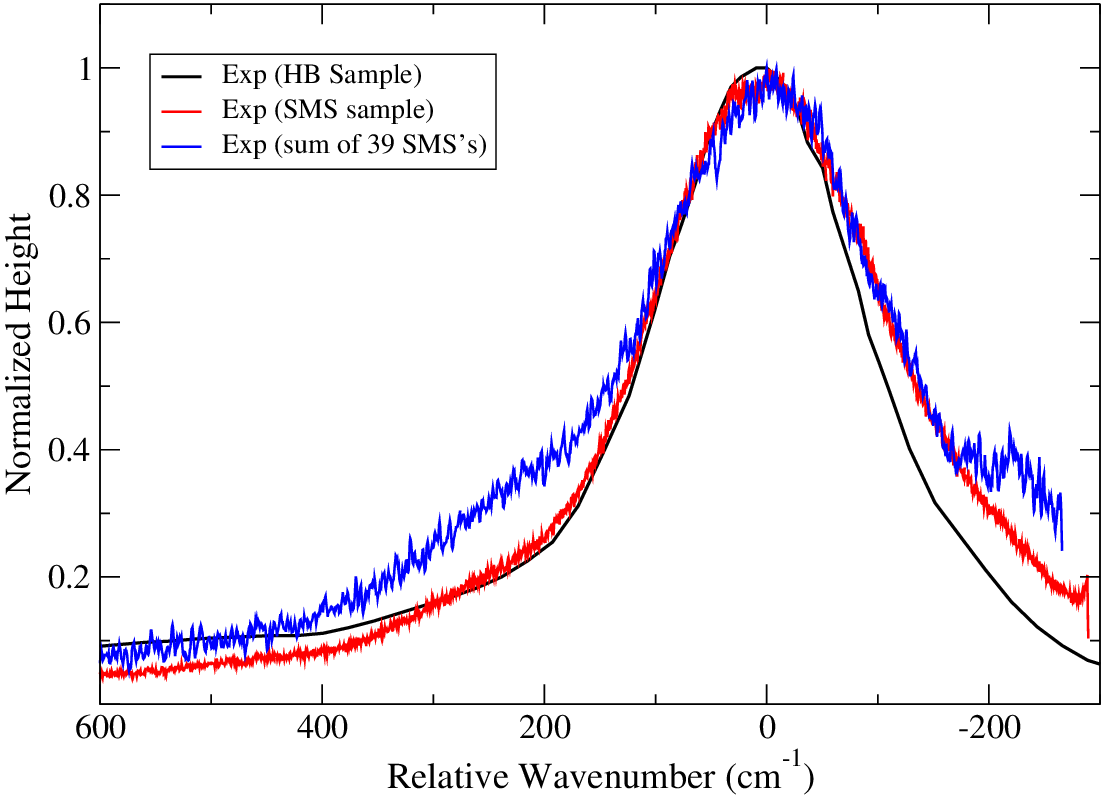} \\
\caption{ Comparison of the ensemble absorption lineshape of B850 obtained from a hole burning sample of LH2 ( the Small group\cite{matsuzaki-jpcb105}) at ${\rm 4\ K}$ and the fluorescence-excitation ensemble lineshape (or sum of 39 SML's) for a sample of LH2 used in the SMS at ${\rm 1.4\ K}$.  }  
\end{figure}

The SM lineshape (SML)\cite{vanojien-science285} of individual B850 unit consists of 2-4 distinctive 
peaks. Out of these,  two major peaks appear consistently, which are indicative of the two exciton states, $|1-\rangle$ and $1+\rangle$, according to their excitation energies and relative polarizations.   Thus, defining ``Gap" for each SML  as the difference between the energies of the two peak maxima, it is reasonable to assume that \cite{vanojien-science285,matsushita-bj80,hofmann-cpl395}  
\be
{\rm Gap} ={\mathcal E}_{1+}-{\mathcal E}_{1-} \  .\label{eq:gap_def}
\ee
Significant number of SML's contain 1-2 additional peaks\cite{vanojien-science285,matsushita-bj80,hofmann-cpl395} in the higher energy side of the two major peaks,  which correspond to higher exciton states.  Because these higher excitonic states are not always observable with good resolution and their assignment ($k=\pm 2$ or $k=\pm 3$) is not always obvious, reliable statistical analysis of these states remains difficult.

With the assignment of Eq. (\ref{eq:gap_def}), van Oijen  {\it et al.}\cite{vanojien-science285} was able to collect the information on the distribution of the splittings between ${\mathcal E}_{1+}$ and ${\mathcal E}_{1-}$.  It turned out that the resulting experimental distribution had much larger average and variance than those predicted from the exciton model based on the X-ray structure.\cite{mcdermott-nature374} In order to explain this discrepancy, van Oijen {\it et al.}\cite{vanojien-science285,matsushita-bj80} suggested elliptic deformation of the LH2 ring.     Later, Hofmann {\it et al.}\cite{hofmann-cpl395} analyzed a more extensive set of single molecule spectroscopy (SMS) data, and came up with an improved model that assumes the elliptic modulation (more precisely, modulation of $k=2$ symmetry) in the excitation energies of chromophores,  which were shown to explain more extensive features of the SMS data.   

Even if the modulation is not physical but rather in the site excitation energy, the assumption of $k=2$ symmetry modulation is somewhat unusual.  Nor has it been supported by other types of spectroscopy.  Does this indicate an unusual sensitivity of the SMS?  Or, is it possible for the sample used in the SMS is qualitatively different from  those used in other spectroscopy?   Comparison of the ensemble lineshapes of  B850 for different samples  is important in answering this question. 

Figure 1 compares two ensemble lineshapes of B850 unit.  One is the fluorescence-excitation ensemble lineshape taken for the sample\cite{hofmann-cpl395} used for the SMS at ${\rm1.4\ K}$.  The other is the conventional absorption lineshape taken by the the Small group\cite{matsuzaki-jpcb105} at ${\rm 4\ K}$ for a sample used for hole burning (HB) spectroscopy.  Extra care was taken to minimize the disorder in this sample.
Though not dramatic, clear difference can be seen between the two ensemble lineshapes, especially on the lower energy side of the peak.   This difference is too big to be attributed to the small temperature differences, and clearly indicates that the two samples have different types or magnitudes of disorder. 
Figure 1 also shows the explicit sum of 39 SML's\cite{hofmann-cpl395} used for statistical analysis, which shows some discrepancies from both ensemble lineshapes.     This indicates that the subset of B850 units employed for statistical analysis of SMS might be biased toward certain types of disorder.

\section{Model of Elliptic Modulation}
The exciton Hamiltonian employed by Hofmann {\it et al.}\cite{hofmann-cpl395} is a simplified one that treats $\alpha$ and $\beta$ BChls equivalently except for the site energy.  Here, we construct the elliptic distortion model based on more realistic exciton Hamiltonian, Eqs. (\ref{eq:h_0}) and (\ref{eq:h_e}). In the notation of present work, the model of $k=2$ site energy modulation suggested by Hofmann {\it et al.}\cite{hofmann-cpl395} can be represented  by the following disorder Hamiltonian:
\ben
\delta H_e&=&\sum_{n=1}^9 \left \{(\xi_{\alpha_n}+E_{mod}\cos(2\cdot 2n\pi/9)-\xi_c) |\alpha_n\rangle\langle \alpha_n| \nonumber \right .\\
&+&\left . (\xi_{\beta_n}+E_{mod}\cos(2\cdot (2n+1)\pi/9)-\xi_c)|\beta_n\rangle\langle \beta_n| \right \} \nonumber \\
\een
where $\xi_{\alpha_n}$ and $\xi_{\beta_n}$ represent Gaussian random variables in the site excitation energies of $\alpha_n$ and $\beta_n$ BChls, and $\xi_c$ is that of the ground electronic state in each complex (so called inter-complex disorder).  $E_{mod}$ is the magnitude of the modulation of $k=2$ symmetry in site excitation energy.  A range of parameters  were tested for this type of disorder and modulation.  We found that the set of parameters\endnote{In this choice, the assumption of bias, $240\ {\rm cm^{-1}}$, in the excitation energy between $\alpha$ and $\beta$ BChls is not appear to be essential, but we kept this number as close to the model of Hofmann {\it et al.}.  For other models, we kept the bias zero in order to minimize the set of model parameters.} shown in the list A of Table I  produce results similar  to those obtained by  Hofmann {\it et al.}\cite{hofmann-cpl395}  Figure 2 shows simulation results, which were obtained by sampling over $1,000,000$ realizations of disorder.\endnote{For reasonable statistics, sampling over about $10,000$ is sufficient, but here we conducted averaging over much larger set in order to minimize ambiguity due to statistical errors.}  The experimental results of SMS are also shown for comparison. 

\begin{table}
\caption{Four models of disorder in the B850 exciton Hamiltonian.  $\epsilon_\alpha$ and $\epsilon_\beta$ represent the average site excitation energies of $\alpha$ and $\beta$ BChls. The symbol $\sigma$ represents the standard deviation of each Gaussian disorder.  Thus, $\sigma_c$ is for the ground state energy (or intercomplex disorder), $\sigma_{\alpha(\beta)}$ is for the $\alpha (\beta)$ site excitation energy, and  $\sigma_E^{k=1}$ and $\sigma_E^{k=2}$ are those for the amplitudes of correlated site energy disorders with $k=1$ and $k=2$, respectively.    $E_{mod}^{k=2}$ is the magnitude of the $k=2$ symmetry modulation of site energies. All the numbers are in the units of ${\rm cm^{-1}}$.\vspace{.1in} }
\begin{tabular}{ccccccc}
\hline
\hline
 &\ \ $\epsilon_\alpha-\epsilon_\beta $ & $\sigma_c $\ \ &\ \ $\sigma_{\alpha(\beta)} $\ \ &\ \ $E_{mod}^{k=2}$\ \ &\ \ $\sigma_E^{k=1}$\ \ &\ \ $\sigma_E^{k=2}$ \\
\hline
\makebox[.3in]{$A$ }& 240 & 50 & 117 & 163 & 0 & 0  \\
\makebox[.3in]{$B$ }& 0   & 30 & 190 & 0   & 0 & 0  \\ 
\makebox[.3in]{$C$}& 0   & 20 & 210 & 0   & 90 & 90   \\
\makebox[.3in]{$D$}& 0 & 30 & 210 & 0& 180 &  0 \\
\hline
\hline
\end{tabular}
\end{table}

In Fig. 2, the agreement between experiment and simulation is reasonable for three distributions -- the distribution of the Gap defined by Eq. (\ref{eq:gap_def}), that of the intensity ratios between the two major peaks, and that of their relative polarization angles.   While these confirm the findings of Hofmann {\it et al.},\cite{hofmann-cpl395} they do not guarantee that the Model A can also explain other information extracted from the SMS data. As an example, we considered the correlation between the intensity ratio and the Gap.  The experimental results are shown in the right bottom panel of Fig. 2.  Red dots represent all the experimental data points obtained from SMS, and the red dashed line represents the lowest order statistical data, the average intensity ratio vs. the Gap.  Although the statistics is not satisfactory, the experimental trend is that the intensity ratio decreases as the Gap increases.  Compared to this is the theoretical value of $\langle I_{1+}/I_{1-}\rangle$ vs. Gap calculated from simulation, which is shown as blue dashed line.  It is clear that the theoretical value is significantly larger than the experimental one and is insensitive to the value of Gap.  

In most SML's, the peak for $|0\rangle$ state is not easy to identify.  One of the reasons for this may be that the $|0\rangle$  peak is merged with the  $|1-\rangle$ peak.  Indeed, the comparison of intensity ratios in the top right panel of Fig. 2 shows that the experimental distribution of the ratios of the higher to lower major peaks agrees better with the distribution of intensity ratios, $I_{1+}/(I_{1-}+I_0)$. Thus, we calculated  $\langle I_{1+}/(I_{1-}+I_{0})\rangle $ as a function of the Gap from the simulation data, which is shown as black solid line.    This average value is closer to the experimental value.  However, the trend with respect to the value of Gap is opposite.  Thus, the experimental trend cannot be well explained yet.

\begin{figure}
\vspace{.4in}
\includegraphics[width=3.in]{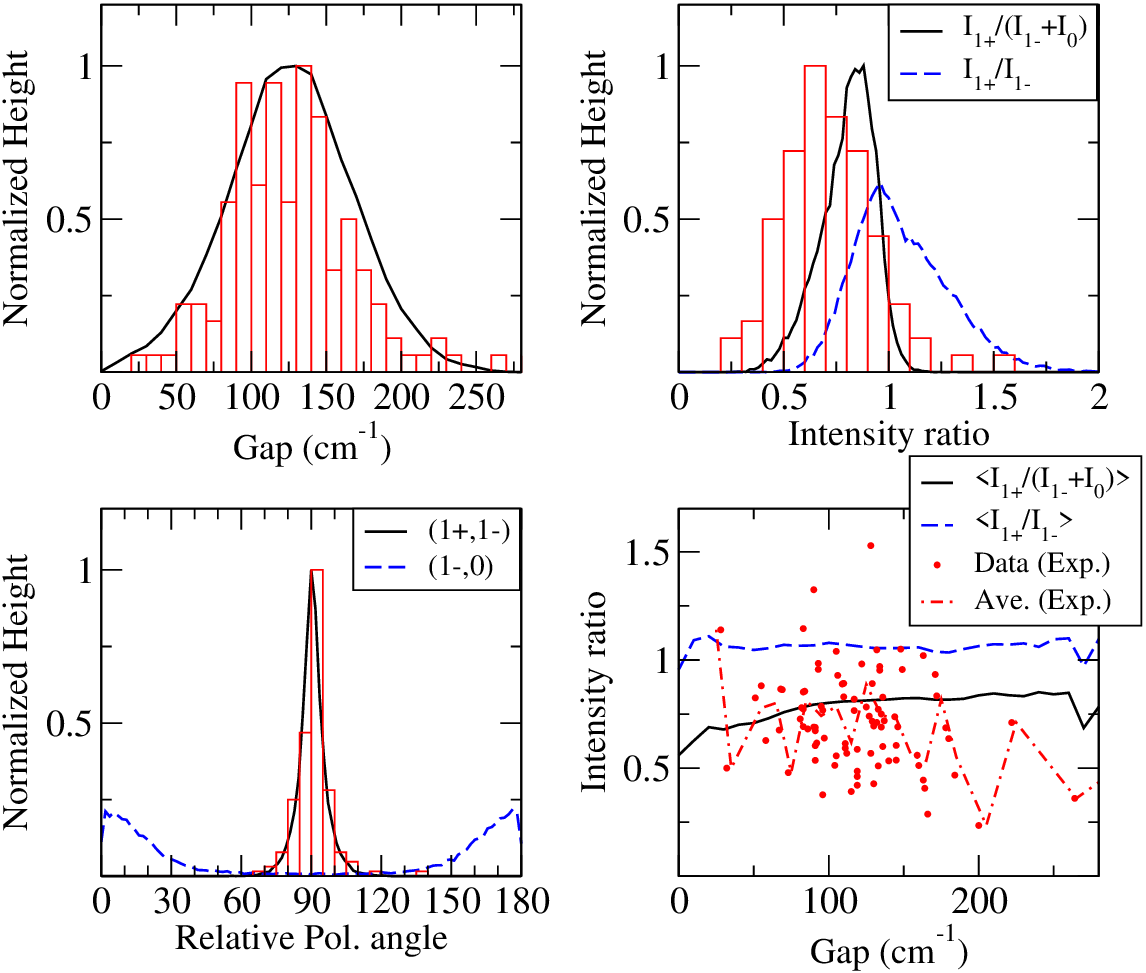} \\
\caption{Simulation results and experimental SMS data for model A in Table 1.  The top left panel shows the distribution of  the gap,  the top right panel shows the distribution of intensity ratios, the bottom left the distribution of relative polarization angles, and the bottom right  the correlation between the intensity ratio and gap.} 
\end{figure}

Another important test of the Model A is to examine whether the theoretical ensemble lineshape based on the model agrees with the experimental ensemble lineshape {\it taken for the same sample}.  Ketelaars {\it et al.}\cite{ketelaars-bpj80} made such comparison, but found significant discrepancy.  They attributed this to the effect of exciton-phonon coupling, which was not considered in their theoretical modeling.\cite{ketelaars-bpj80} This motivates the need to calculate more reliable ensemble lineshape including the effect of electron-phonon coupling.

\section{Exciton Bath Model and Ensemble Lineshape}

The effects of coupling between electronic excitations and protein phonon modes in B850 can be accounted for by a total Hamiltonian $H=H_e+H_{eb}+H_b$, where $H_b$ represents the phonon bath and $H_{eb}$ the exciton-bath coupling.  It  is assumed that the bath Hamiltonian can be modeled by an infinite number of harmonic oscillators with  a well-defined spectral density and that  $H_{eb}$ is linear in the bath coordinate.\cite{jang-jcp118-2}

At the level of a 2nd order Quantum Master Equation (QME), the ideal line shape\cite{jang-jcp118-2} of a single B850 complex (the lineshape of a single B850 in the absence of quasistatic disorder) can be calculated by the following expression:
\be
I(\omega)\approx -\frac{1}{3\pi}\sum_{\bf \hat e=\hat x, \hat y, \hat z} {\rm Im}\ Tr_e\left\{\frac{{\bf \hat e}\cdot|{\bf D}\rangle\langle {\bf D}|\cdot{\bf \hat e}}{\omega+(\epsilon_g-H_e)/\hbar+i\hat {\mathcal K}(\omega)}\right\} \ , \label{eq:ie}
\ee
where  ``${\rm Im}$" means taking imaginary part of the complex function, $Tr_e$ represents trace operation of the electronic degrees of freedom only, $\epsilon_g$ is the electronic ground state energy of each BChl,  ${\bf \hat e}$ is the polarization of the 
radiation, $|{\bf D}\rangle=\sum_{n=1}^9 \sum_{s=\alpha}^\beta \mbox{\boldmath $\mu$}_{s_n}|s_n\rangle $ is the superposition of exciton states weighted by transition dipole vectors, and  
\be
\hat {\mathcal K} (\omega)=\sum_{n=1}^N\sum_{p=0}^{8}\sum_{b=l}^u \hat \kappa (\omega-{\mathcal E}_{b,p}/\hbar)\sum_{s=\alpha}^{\beta}|C^{s,b}_{n,p}|^2 |s_n\rangle \langle s_n| \ .  \label{eq:hat_K}
\ee
In the above expression, $\hat\kappa (\omega)$ is the Fourier transform of the time correlation function  of phonons coupled to a single BChl and is expressed as 
\ben
&&\hat \kappa (\omega)\equiv\int_0^\infty dt\ e^{i\omega t}\ \int_0^\infty d\omega {\mathcal D} (\omega)\nonumber \\ && \hspace{.4in}\times \left \{\coth(\frac{\hbar\omega}{2k_BT})\cos(\omega t)-i \sin(\omega t)\right\}\ ,
\een 
where ${\mathcal D}(\omega)$ is the spectral density\cite{renger-jcp116} of the phonon modes coupled to the $Q_y$ excitation of each BChl and has been shown to be\cite{jang-jpcb111} well modeled by the following function:  ${\mathcal D}(\omega)=0.22\omega e^{-\omega/\omega_{c1}}+0.78(\omega^2/\omega_{c2})e^{-\omega/\omega_{c2}}+0.31(\omega^3/\omega_{c3}^2)e^{-\omega/\omega_{c3}}$ with $\omega_{c1}=170 {\rm \ cm^{-1}}$, $\omega_{c2}=34{\rm \ cm^{-1}}$, and $\omega_{c3}=69{\rm \ cm^{-1}}$.  It was assumed that $k_BT=10{\rm \ cm^{-1}}$, which serves as the reasonable cryogenic temperature limit.  In Eq. (\ref{eq:hat_K}), $C^{s,b}_{n,p}$ is the matrix element of the transformation that relates  
$\{|\alpha_n\rangle, |\beta_n \rangle\} $ with $\{ |\psi_{l,p}\rangle, |\psi_{b,p}\rangle\} $ as follows: $|s_n\rangle=\sum_{p=0}^{N-1}\left\{C_{n,p}^{s,l} |\psi_{l,p}\rangle + C_{n,p}^{s,u}|\psi_{u,p}\rangle \right\}$ for $s=\alpha,\beta$.

The calculation of $I(\omega)$ for each $\omega$ involves that of $\omega +(\epsilon_g-H_e)/\hbar +i\hat {\mathcal K} (\omega)$, which has the same dimension as the exciton Hamiltonian $H_e$, followed by numerical inversion.  The whole calculation in the spectral region of B850 for each complex can be conducted in tens of seconds.  This enables a theoretical ensemble lineshape  for any model of disorder to be calculated by averaging Eq. (\ref{eq:ie}) over many realizations of disorder.   Averaging over about 40,000 realizations of the disorder took less than a day in a typical single processor workstation.\endnote{The resulting theoretical lineshape has much more noise on the red side of the peak than the blue side.  This is because the major contribution to the red side comes from the peaks of $|1-\rangle$ and $|0\rangle$, which have narrower line widths.  }   

In calculating the ensemble lineshape,  Gaussian disorder in the ground state energy (or intercomplex disorder) with standard deviation of $\sigma_c=50\ {\rm cm^{-1}}$ was assumed, which is somewhat large but is a minimum value necessary to make the theoretical ensemble line shape single-peaked (The whole set of parameters including this parameter is represented as Model $A$ in Table I.). The result is compared with the experimental lineshape in Fig. 3. Significant difference can be seen between theoretical and experimental ensemble lineshapes.  The theoretical lineshape is broader and decreases much more steeply on the red side than the experimental one.  Recent theoretical studies\cite{schroder-jcp124,chen-jcp131} suggest inaccuracy of the 2nd order time non-local QME approach at room temperature.  However, this effect is expected to be small in the cryogenic temperature limit considered here.  In addition, the multiphonon effect, which serves as the major source of the theoretical error, is not likely to explain both the discrepancy in the red side of the peak and the width of the lineshape being broader than the experimental one.  

\begin{figure}
\vspace{.4in}
\includegraphics[width=3. in]{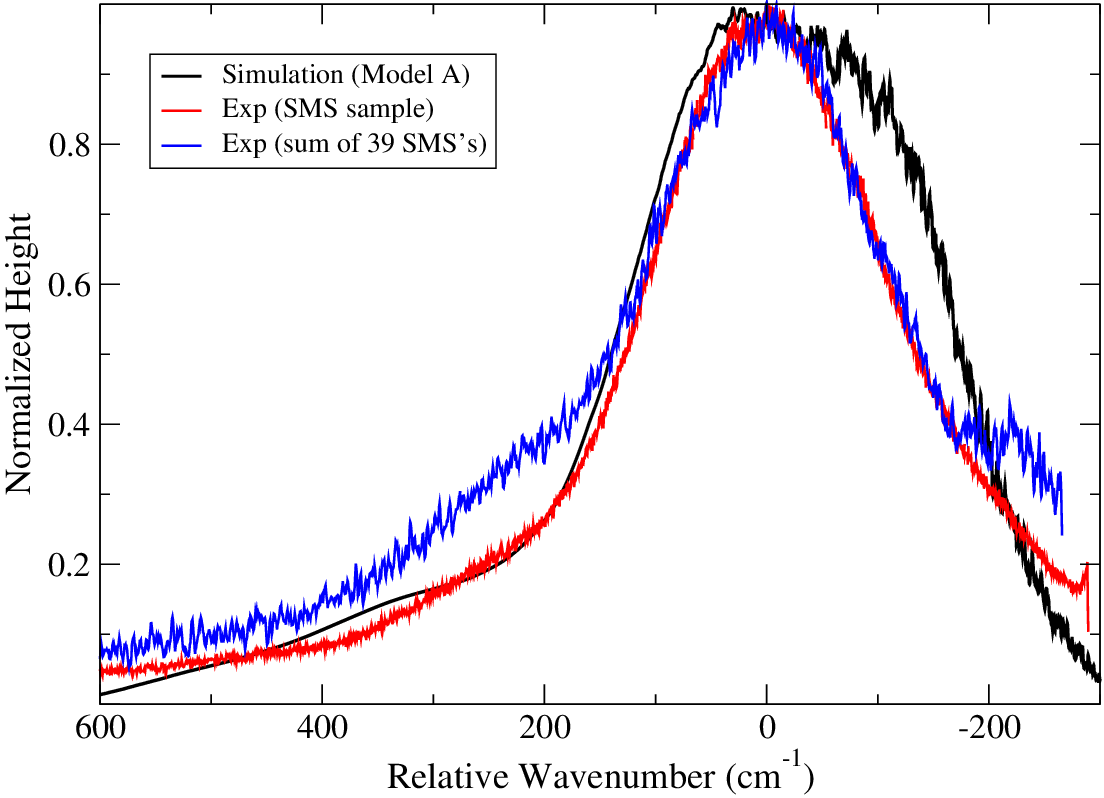}\\
\caption{ Comparison of theoretical ensemble lineshape (for Model A) with experimental ensemble lineshape and the sum of SML's. } 
\end{figure}

The discrepancy between the theoretical and experimental ensemble line shapes in Fig. 3 is quite substantial when compared with the performance of other alternative models.   For example, the ensemble line shape taken for the HB sample by the Small group\cite{matsuzaki-jpcb105} can be modeled quite well by assuming simple Gaussian disorder only.  The parameters of Model $B$ in Table I is one such example providing good fit of experimental ensemble lineshape.  Figure 4 compares the theoretical ensemble lineshape calculated by averaging Eq. (\ref{eq:ie}) over $40,000$ realizations of Model $B$ disorder with the experimental ensemble lineshape. The agreement between theoretical and experimental lineshapes is excellent except in the far blue side of the peak where possible multiphonon effects and some contribution of B800 may exist. 

\begin{figure}
\vspace{.4in}
\includegraphics[width=3. in]{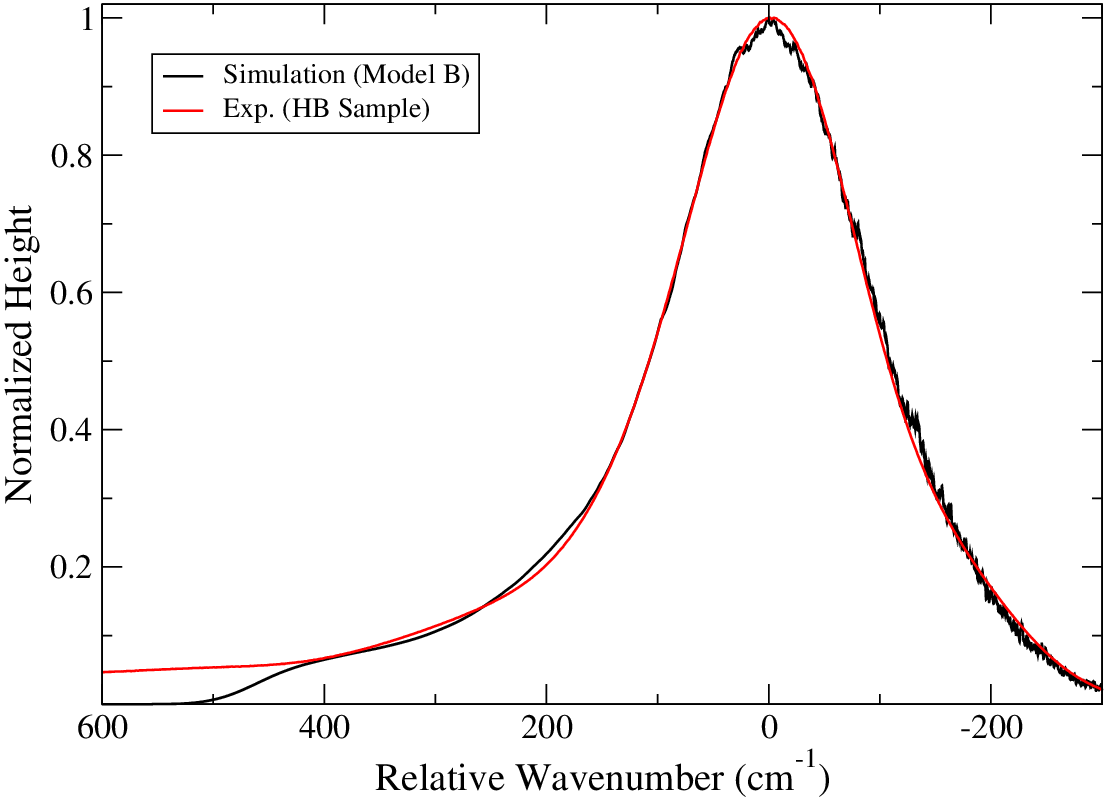}\\
\caption{ Comparison of theoretical ensemble lineshape (for Model B) with experimental ensemble lineshape for a sampled used by the Small group in the HB spectroscopy. The fall of theoretical lineshape at a value a little above ${\rm 400 \ cm^{-1}}$ reflects the fact the ideal lineshape of each B850 unit before averaging over disorder was calculated only up to this value.  } 
\end{figure}

\begin{figure}
\includegraphics[width=3. in]{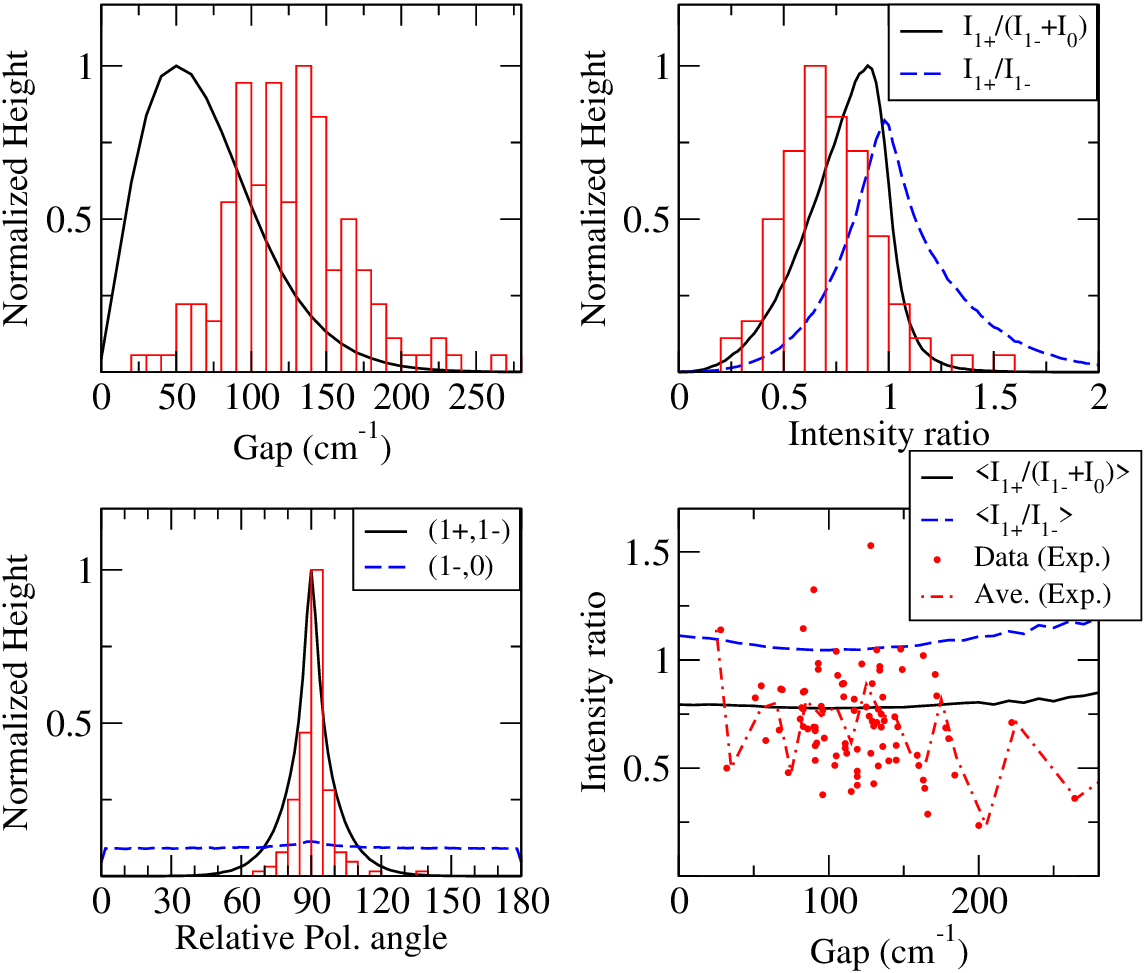}
\caption{Comparison of the simulation of SMS data for Model B with experimental results. Other details are the same as Fig. 2.}
\end{figure}

For the Model $B$, SMS data were calculated by sampling the exciton Hamiltonian over $1000,000$ realizations of the disorder, which are compared against experimental data in Fig. 5.  As was noted before,\cite{vanojien-science285} the theoretical distribution of the Gap has much smaller average than the experimental one. The correlation between the intensity ratio and the Gap appears to fit the experimental data better than Model $A$, but the agreement is not satisfactory yet.   Considering that different models of disorder are needed for the samples of SMS and HB, these are not unexpected.  However, this raises the following question naturally.  Is there a model of disorder or distortion which reproduces both the {\it ensemble lineshape} and {\it other statistical data of the SMS} for the same sample of LH2?  Our analysis above shows that  the model A, which assumes uniform elliptic distortion or $k=2$ symmetry modulation in the site excitation energy, does not serve that role.    Thus, it is necessary to consider other alternative models of disorder or distortion to answer this question.

It is plausible to assume that the disorder with long range spatial correlation is more dominant than that with short length scale for the SMS sample of LH2's dispersed on a disordered polymeric surface at the cryogenic temperature limit because the potential barrier associated with the former is expected to be larger than that of the latter.  
However, there is no specific reason why such long range disorder has only $k=2$ symmetry and uniform amplitude.  Thus, we here introduce a new model of disorder represented by the following disorder Hamiltonian: 
\ben
\delta H_e &=& \sum_{n=1}^9 \left\{(\xi_{\alpha_n}+\xi_1\cos(2n\pi/9+\varphi_{1\alpha})\right . \nonumber \\
&&\hspace{.2in}+ \xi_2\cos(4n\pi/9+\varphi_{2\alpha})-\xi_c)|\alpha_n\rangle\langle \alpha_n|\nonumber \\
&&+(\xi_{\beta_n}+\xi_1\cos(2n\pi/9+\varphi_{1\beta}) \nonumber \\
&&\left .\hspace{.2in}+ \xi_2\cos(4n\pi/9+\varphi_{2\beta})-\xi_c)|\beta_n\rangle\langle \beta_n|\right\} \label{eq:delta_h_en}
\een
where $\xi_{\alpha_n}$ and $\xi_{\beta_n}$ are the same random Gaussian disorder of site excitation energy as introduced before, $\xi_1$ and $\xi_2$ are random Gaussian variables representing the amplitudes of correlated disorder in site excitation energies with $k=1$ and $k=2$ symmetries.  The standard deviations of these two types of disorder are respectively denoted as $\sigma_{E}^{k=1}$ and $\sigma_E^{k=2}$.  In general, four phase variables, $\varphi_{1\alpha}$, $\varphi_{2\alpha}$, $\varphi_{1\beta}$, and $\varphi_{2\beta}$, can be assumed to be arbitrary and  these are sampled from uniform random variables.

We have tested different sets of parameters and compared the  theoretical ensemble lineshape with the experimental ensemble lineshape of the SMS sample.  It was possible to come up with many similar sets of model parameters  that can reproduce the experimental ensemble lineshape with reasonable accuracy.  The Model $C$ in Table I is a representative of those models of disorder.   
Figure 6 compares the theoretical ensemble lineshape based on this model with the experimental lineshapes.  The agreement with experiment is much better than that for Model $A$ shown in Fig.  3.  We also calculated the SMS data, and the simulation results are compared with experimental data in Fig. 7.   While other statistical distributions from the SMS are now in reasonable agreement with the experimental data, significant discrepancy can be seen in the distribution of the Gap.

\begin{figure}
\vspace{.4 in}
\includegraphics[width=3. in]{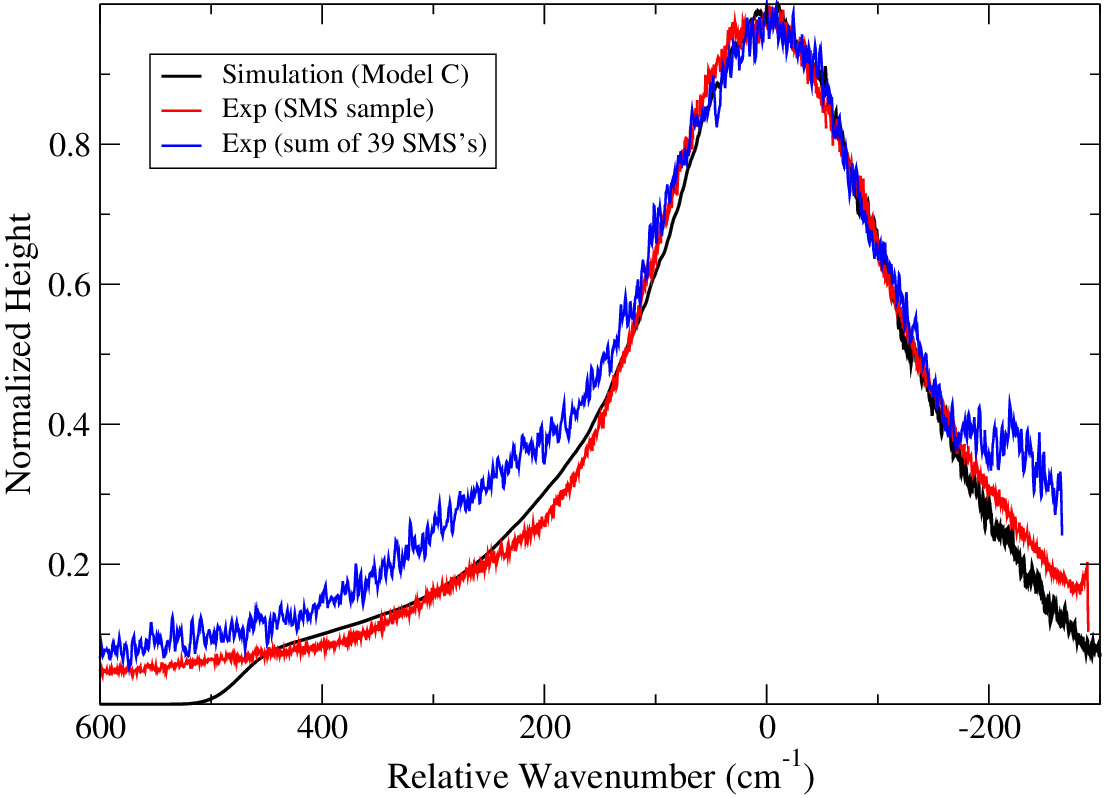}\\
\caption{Comparison of experimental and theoretical ensemble absorption lineshape for Model C.}
\end{figure}

\begin{figure}
\includegraphics[width=3. in]{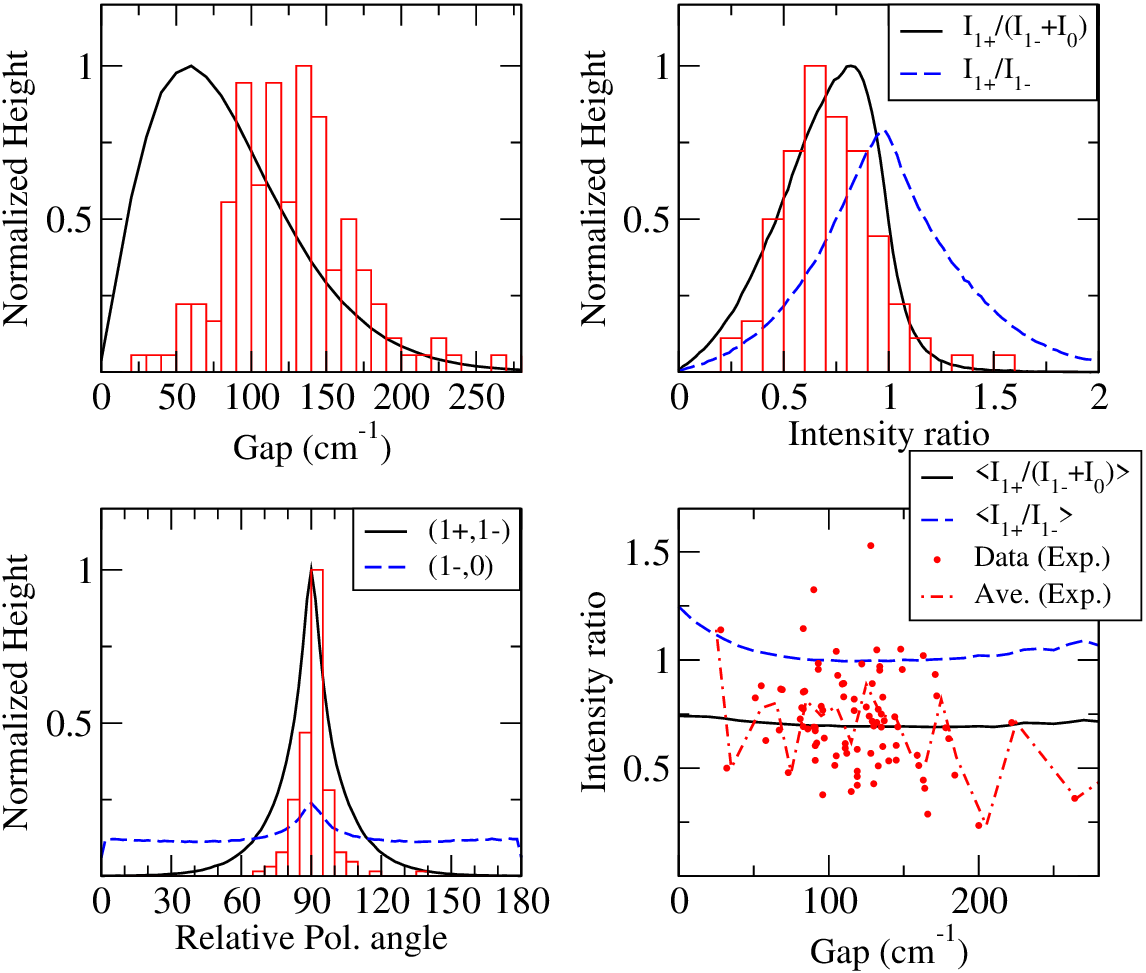}\\
\caption{Comparison of the simulation of SMS data for Model C with experimental results. Other details are the same as Fig. 2. }
\end{figure}

Within the form of the disorder Hamiltonian, Eq. (\ref{eq:delta_h_en}), we also tested another scenario where there is only $k=1$ symmetry correlation in the site energy disorder  and  the random Gaussian disorder ($\xi_2=0$).   Based on the test of a range of disorder parameters, we identified a model that reproduce the sum of SML's (not the ensemble lineshape) as closely as possible.  This is denoted as  the Model $D$ in Table I.  The resulting ensemble lineshape is compared with experimental lineshapes in Figure 8.  This suggests the possibility that the set of LH2's selected for SMS tend to have larger $k=1$ symmetry disorder.  This is understandable considering that the $k=1$ symmetry enhances the oscillator strength of the lowest exciton state $|0\rangle$.   We also calculated the various statistical distributions that can be obtained from the SMS data. The results are compared with experimental data in  Fig. 9.   As in the case of Model C, the agreement in the Gap distribution is not satisfactory while others show good agreement with the experimental data.

 \begin{figure}
 \vspace{.4in}
\includegraphics[width=3. in]{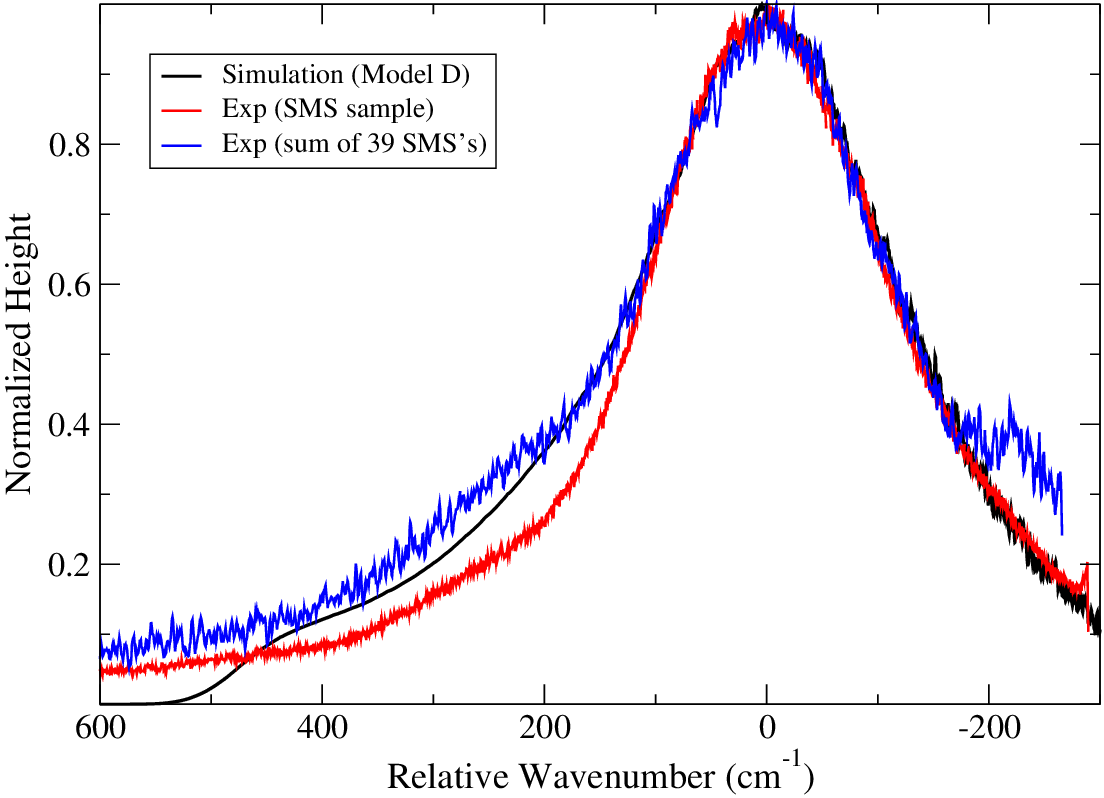}\\
\caption{Comparison of experimental and theoretical ensemble absorption lineshape for Model D.}
\end{figure}

\begin{figure}
\includegraphics[width=3. in]{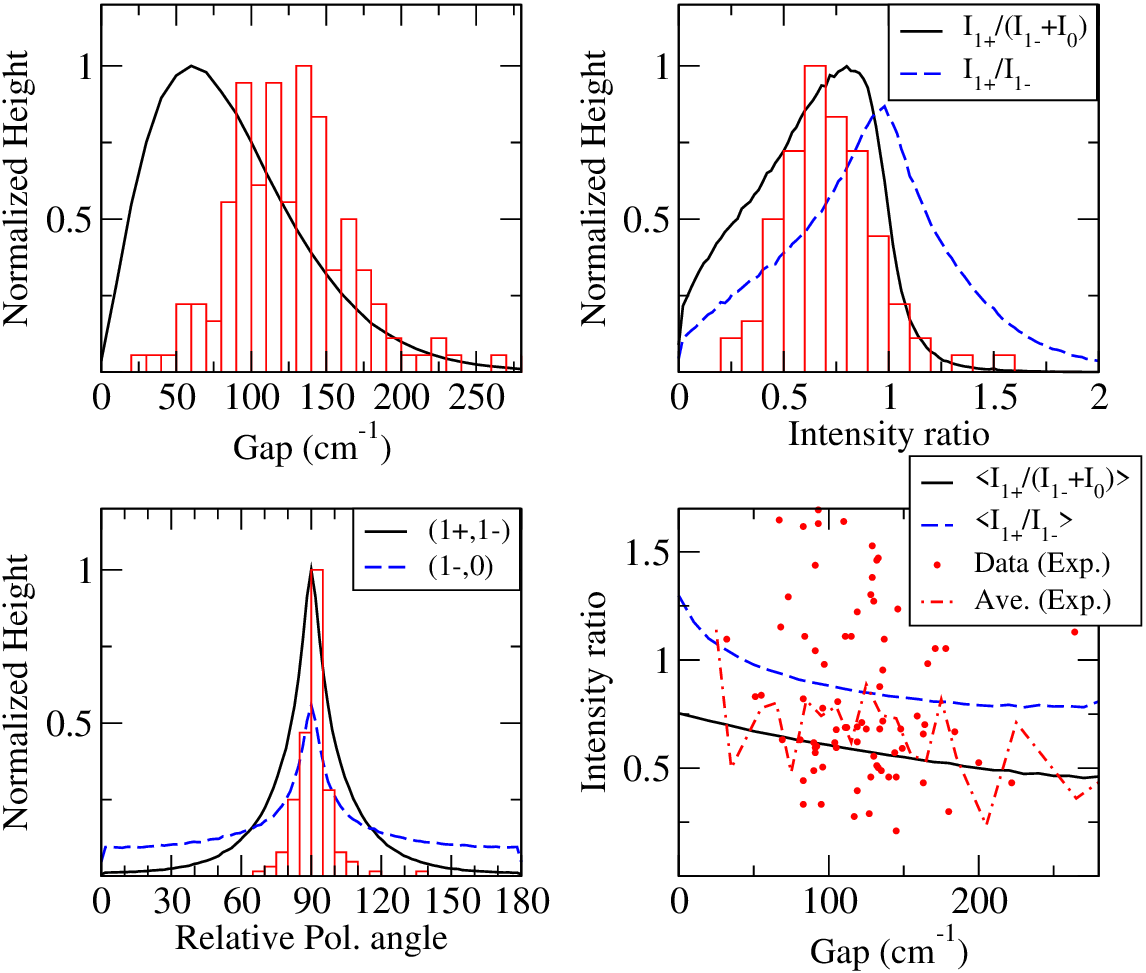}\\
\caption{Comparison of the simulation of SMS data for Model D with experimental results. Other details are the same as Fig. 2.}
\end{figure}

\section{New interpretation of SMS data}

The results in Figs. 7 and 9  show that both models $C$ and $D$ can be viewed as reasonable representations of the SMS data except for the Gap distribution.  One may attribute this to the poor statistics of SMS but it is not likely that only the Gap distribution is subject to such error.  Thus, clarification of this issue is necessary. In fact, a careful examination of the analysis made so far suggests an alternative explanation.

In the top right panels of all the figures reporting the simulation of the SMS data for all the models, it is clear that the distribution of experimental intensity ratios can be modeled better by $I_{1+}/(I_{1-}+I_0)$ than by $I_{1+}/I_{1-}$.  This suggests that the lower major peak in SML contain both contributions from $|1-\rangle$ and $|0\rangle$ states.  Then, it is not consistent to assume that the $|0\rangle$ state does not affect the peak position of the lower peak at all.  If we assume the possibility of dynamical averaging between $|1-\rangle$ and $|0\rangle$ during the time scale of SMS, the position of the lower peak should be approximated as an intensity weighted average of ${\mathcal E}_0$ and ${\mathcal E}_{1-}$.  Under this assumption, the Gap measured from the SMS data may indeed be the following quantity:\endnote{To the best of knowledge, the possibility that the contribution of $k=0$ state may be responsible for large gap distribution was raised for the first time by van Grondelle, but systematic investigation of this issue in any publication is not available.}   
\be
{\rm Gap}={\mathcal E}_{1+}-\frac{I_0 {\mathcal E}_0+I_{1-}{\mathcal E}_{1-}}{I_0+I_{1-}} \ . \label{eq:gap_dist}
\ee  

Figure 10 shows the resulting  Gap distributions for all the four models of disorder, and Fig. 11 the new intensity-Gap correlations.  It is important to note that the new theoretical distributions for Models C and D now agree quite well with experimental data.   On the other hand, the Model A now overestimates and the Model B still underestimates the Gap distributions.  Thus, with the new assignment of the Gap, Eq. (\ref{eq:gap_dist}), either Model C or D can provide satisfactory description of  the full SMS data.

\begin{figure}
\vspace{.4in}
\includegraphics[width=3. in]{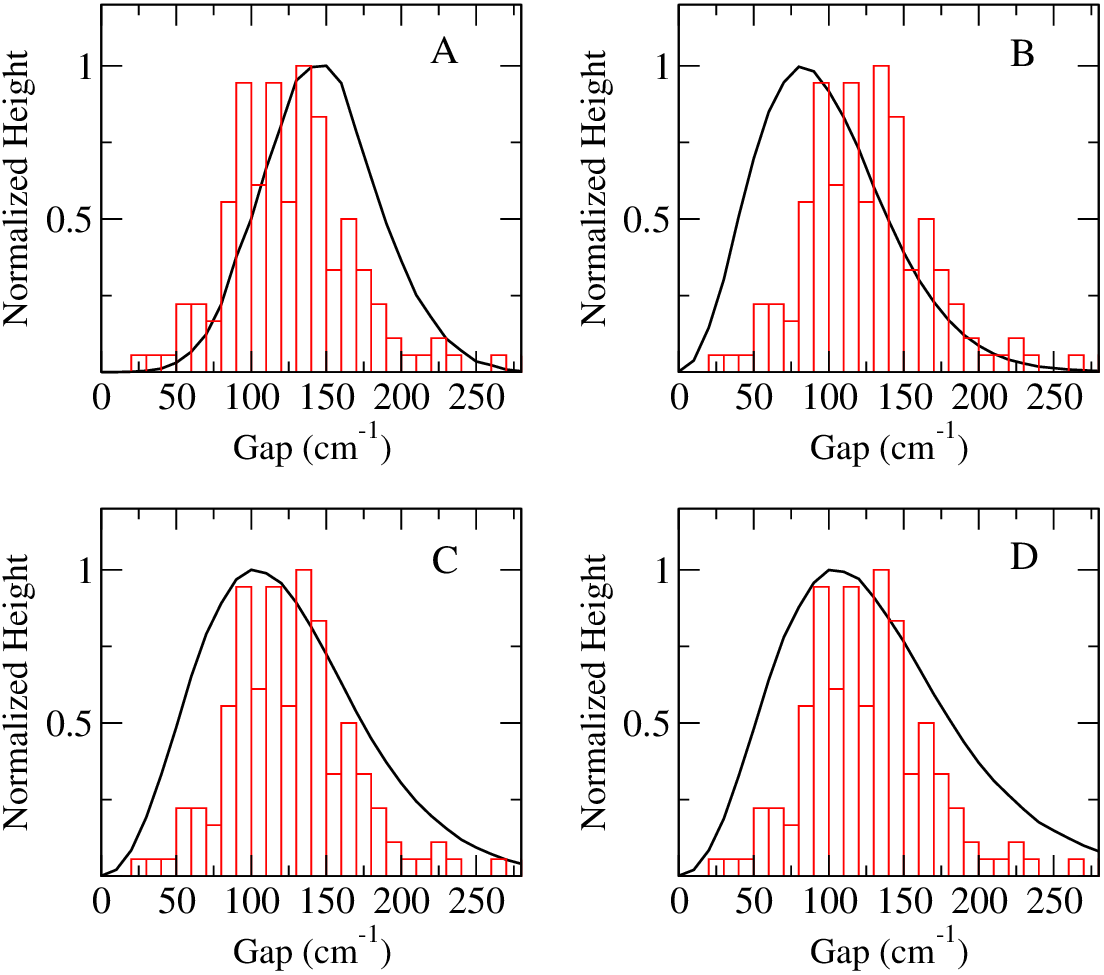}\\
\caption{Comparison of new Gap distributions for four models of disorder in Table I.  Solid lines are simulation results and the histograms are experimental data.  }  
\end{figure}

\begin{figure}
\vspace{.4in}
\includegraphics[width=3. in]{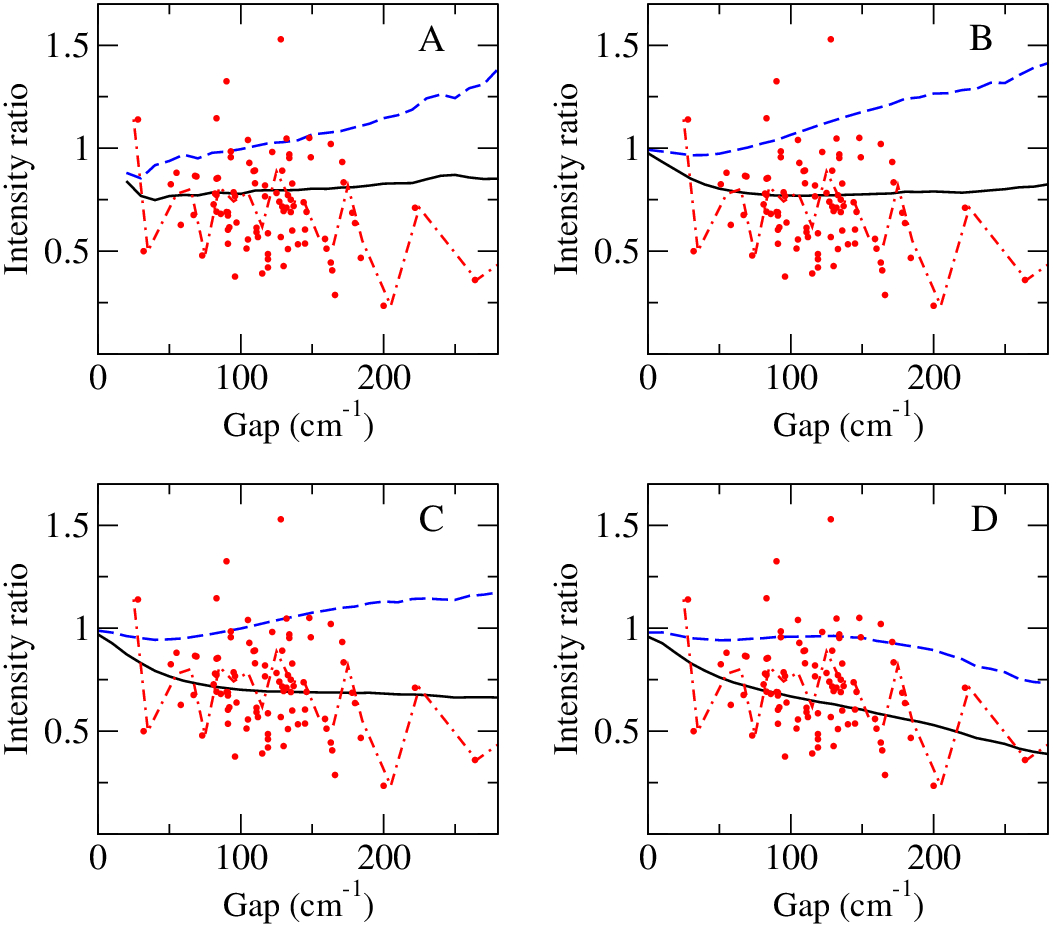}\\
\caption{Comparison of correlation between intensity ratio and the new Gap for four different models of disorder. distribution with two simulation results. The convention different lines and dots are the same as the right bottom panel of Fig. 2.}  
\end{figure}

\section{Concluding Remarks}
In this work, we re-examined the interpretation of low temperature SMS data for the B850 unit of LH2. The issue addressed here was whether there is $k=2$ symmetry modulation in the site excitation energy of LH2. While this model was capable of explaining most SMS data, it has not been clear whether the same assumption can explain other ensemble or subensemble spectroscopic data.   

Based on a more detailed analysis of the SMS data and theoretical modeling of ensemble lineshapes, we came to the conclusion that the assumption of $k=2$ symmetry modulation cannot explain the ensemble lineshape nor the correlation between intensity ratios of the two major peaks and the Gap. 
In order to explain these discrepancies, we tested alternative models of disorder, Model C, which has  disorder with both $k=1$ and $k=2$ symmetries, and Model D, which has disorder with $k=1$ symmetry.  With the use of new interpretation of the Gap, Eq. (\ref{eq:gap_dist}), we found that Model C can explain all the SMS data considered and the ensemble lineshape taken for the same sample.  On the other hand, we found that Model D can explain all the SMS data and the sum of all the 39 SMS lineshapes.     

As can be seen from the comparison with Model B, which reproduces the ensemble lineshape of the Small group, the presence of additional disorder with $k=1$ and $k=2$ symmetry seems not true for any sample.  Rather, it is characteristic of the sample used for the SMS,  although it does not appear to be caused by a specific method of sample preparation.\cite{richter-bj93}   In addition, the fact that the sum of SML's can be explained by the model D suggests that the selection of complexes tend to bias toward those with larger $k=1$ symmetry  disorder and thus with larger fluorescence.  However,  comparison of Figs. 7 and 8 show that such bias is not the reason for unusual characteristics of the SMS data.  

The types of correlation suggested in the present work are static and long range, and are likely\cite{abramavicius-pccp6} due to the quenched deformation of LH2 from its perfect circular symmetry at low temperature.  Thus, they are different from dynamical and short range correlation investigated in recent simulation.\cite{olbrich-jpcb114}  Considering the flexibility of proteins constituting the LH2 and the fact that even small distance changes can result in substantial changes in excitation energies of typical molecular systems, the amount of correlated energetic disorder seen for the sample of SMS is in fact very small, which demonstrates resilience of LH2 in protecting the excitons.   

The lineshape theory employed here is more suitable for low temperature environments, and different approaches\cite{schroder-jcp124,chen-jcp131,novoderezhkin-bj90} have been shown to be successful for the modeling of other types of spectroscopy. For example, hierarchical equation of motion approach\cite{schroder-jcp124,chen-jcp131} works well for the modeling of nonlinear spectroscopy at high temperature and modified Redfield approach was shown to be successful  for modeling single molecule emission spectroscopy.\cite{novoderezhkin-bj90}  From a theoretical point of view, developing a more satisfactory approach that can encompass broader temperature regime and dynamical regime is important for comprehensive understanding of  exciton dynamics in LH2.

\acknowledgments
This research was supported by the Office of Basic Energy Sciences, Department of Energy (Grant No. DE-SC0001393).  SJ also acknowledges partial support from the  National Science Foundation CAREER award (Grant No. CHE-0846899) and the Camille Dreyfus Teacher Scholar Award.  RK and JK thank the German Science Foundation for financial support (DFG,  KO 1359/16 and GRK 1640).

\end{document}